\documentclass[a4paper,11pt]{article}
\usepackage{pos}
\usepackage{MnSymbol}
\newcommand{\Eq}[1]{Eq.~(\ref{#1})}

\newcommand{\Eqs}[2]{Eqs.~(\ref{#1}-\ref{#2})}
\newcommand{\Fig}[1]{Fig.~\ref{#1}}

\title{Toward Exploring Phase Diagrams of Gauge Theories on Quantum Computers with Thermal Pure Quantum States}
\ShortTitle{Phase Diagrams of Gauge Theories with Thermal Pure Quantum States}

\author[a,b]{Zohreh Davoudi}
\author[c]{Niklas Mueller}
\author*[a,b]{Connor Powers}

\affiliation[a]{Maryland Center for Fundamental Physics and Department of Physics, University of Maryland, College~Park, MD~20740, USA}

\affiliation[b]{Institute for Robust Quantum Simulation, University of Maryland, College~Park, Maryland~20742, USA}

\affiliation[c]{InQubator for Quantum Simulation (IQuS), Department of Physics, University of Washington, Seattle, WA~98195, USA.}

\emailAdd{cdpowers@umd.edu}

\abstract{Aiming at evading the notorious sign problem in classical Monte-Carlo approaches to lattice quantum chromodynamics, we present an approach for quantum computing finite-temperature lattice gauge theories at non-zero density. Based on the thermal pure-quantum-state formalism of statistical mechanics when extended to gauge-theory systems, our approach~\cite{davoudi2022toward} allows for sign-problem-free quantum computations of thermal expectation values and non-equal time correlation functions. By taking a simple lattice gauge theory for which classical benchmarks are possible, namely $\mathbb{Z}_2$ lattice gauge theory in 1+1 dimensions at finite chemical potential, we discuss resource requirements and robustness to algorithmic and hardware imperfections for near-term quantum-hardware realizations.}
\FullConference{%
The 39th International Symposium on Lattice Field Theory,\\8-13 August 2022\\
Hörsaalzentrum Poppelsdorf, Bonn, Germany
}

\begin{document}
\maketitle

\section{Introduction}
\noindent
Sign problems plaguing classical Monte-Carlo sampling techniques have precluded quantitative understanding
of the large quark-chemical-potential regime of quantum chromodynamics (QCD), relevant e.g., for relativistic heavy-ion collisions
or neutron and quark stars.
Techniques to overcome the QCD sign problem include reweighting,
Majorana
and Meron Cluster algorithms,
stochastic quantization and complex Langevin dynamics,
Taylor expansion,
analytic continuation,
and path deformation and complexification,
see Refs.~\cite{Nagata:2021ugx,Alexandru:2020wrj} for recent reviews.

In contrast, quantum-computation and simulation techniques do not suffer from sign problems. They offer a promising route toward the inaccessible regime of the QCD phase diagram by directly quantum simulating lattice gauge theories (LGTs), see, e.g., Refs.~\cite{davoudi2022quantum,banuls2020simulating,klco2022standard,martinez2016real,klco2018quantum,nguyen2022digital,mueller2022quantum,de2021quantum,davoudi2022toward,lamm2019general,klco20202,ciavarella2021trailhead,atas2022real}. 
However, thermal i.e. mixed, as opposed to pure, quantum states are na\"ively `unnatural' for quantum computers, making simulations of thermal systems an extensively researched field, which are addressed by a variety of techniques, see e.g., Refs.~\cite{lu2021algorithms,motta2020determining,chowdhury2016quantum,schuckert2022probing,bassman2021computing}.

One promising  route to quantum computing thermal systems is the thermal pure-quantum-(TPQ-) state formulation of statistical mechanics~\cite{sugiura2013canonical}. While originally developed without quantum technology in mind, this ansatz offers a promising route to simulating quantum systems at finite temperature and chemical potential, enabling estimations of thermal expectation values of a large class of observables from only a single properly prepared pure state in the thermodynamic limit~\cite{coopmans2022predicting,powers2021exploring}. Canonical TPQ states are obtained from a Haar-random state evolved in imaginary time~\cite{sugiura2013canonical},
\begin{equation}
\label{eq:defTPQ}
    |\beta,N \rangle \equiv {e^{-\frac{\beta }{2}H}}{}|\psi_R\rangle\,.
\end{equation}
Here,  $\beta$ is the  inverse temperature, $N$ the system size, $H$ the Hamiltonian of the system, and $| \psi_R \rangle $ is a (pseudo-) Haar-random state of the underlying Hilbert space $\mathcal{H}$. One obtains thermal expectation values of low-degree polynomials of local operators through the stochastic average over $r$ TPQ realizations (denoted by $\llangle \cdots \rrangle_r$):
\begin{align}\label{eq:TPQobs}
    \langle O \rangle_\beta \approx \frac{\llangle  \, \langle \beta,N | O | \beta,N \rangle \, \rrangle_r}{\llangle  \, \langle \beta,N | \beta,N \rangle \, \rrangle_r}\,,
\end{align}
where in the thermodynamic limit, i.e., $N\rightarrow \infty$, only a single TPQ state suffices~\cite{sugiura2013canonical}.

\section{Physical thermal pure quantum states}
\noindent
The physical Hilbert space of gauge theories is often a small subset of a larger Hilbert space, i.e., $\mathcal{H}_G \subset \mathcal{H}$, where $| \psi \rangle \in \mathcal{H}_G$ is physical only if $G_n | \psi \rangle = g^{\rm phys} | \psi \rangle $. Here, $G_n$ is the (local) Gauss's law operator that commutes with the Hamiltonian, $g^{\rm phys}$ is the eigenvalue associated with the physical sector ($g^{\rm phys}=1$ for the case of $\mathbb{Z}_2$ LGT considered later), and $n$ labels a lattice site. Because of these local constraints, \Eqs{eq:defTPQ}{eq:TPQobs} do not apply to LGTs unless $|\psi_R\rangle$ is also restricted to the physical Hilbert space, or a penalty term is incorporated in the (P)TPQ-state construction, i.e.,
\begin{align}
\label{eq:physdefTPQ}
|\beta,N \rangle^{\rm phys} \equiv {e^{-\frac{\beta }{2}\widetilde{H}}}|\Psi_R\rangle\,,
\end{align}
where $\widetilde{H}\equiv H+ \sum_n f(G_n)$, and $f(G_n)$ is chosen such that unphysical components are penalized during imaginary-time evolution~\cite{halimeh2022gauge,mathew2022protecting}. This latter approach is what is investigated in this work.

\begin{figure}[t!]
    \centering
    \includegraphics[scale=0.645]{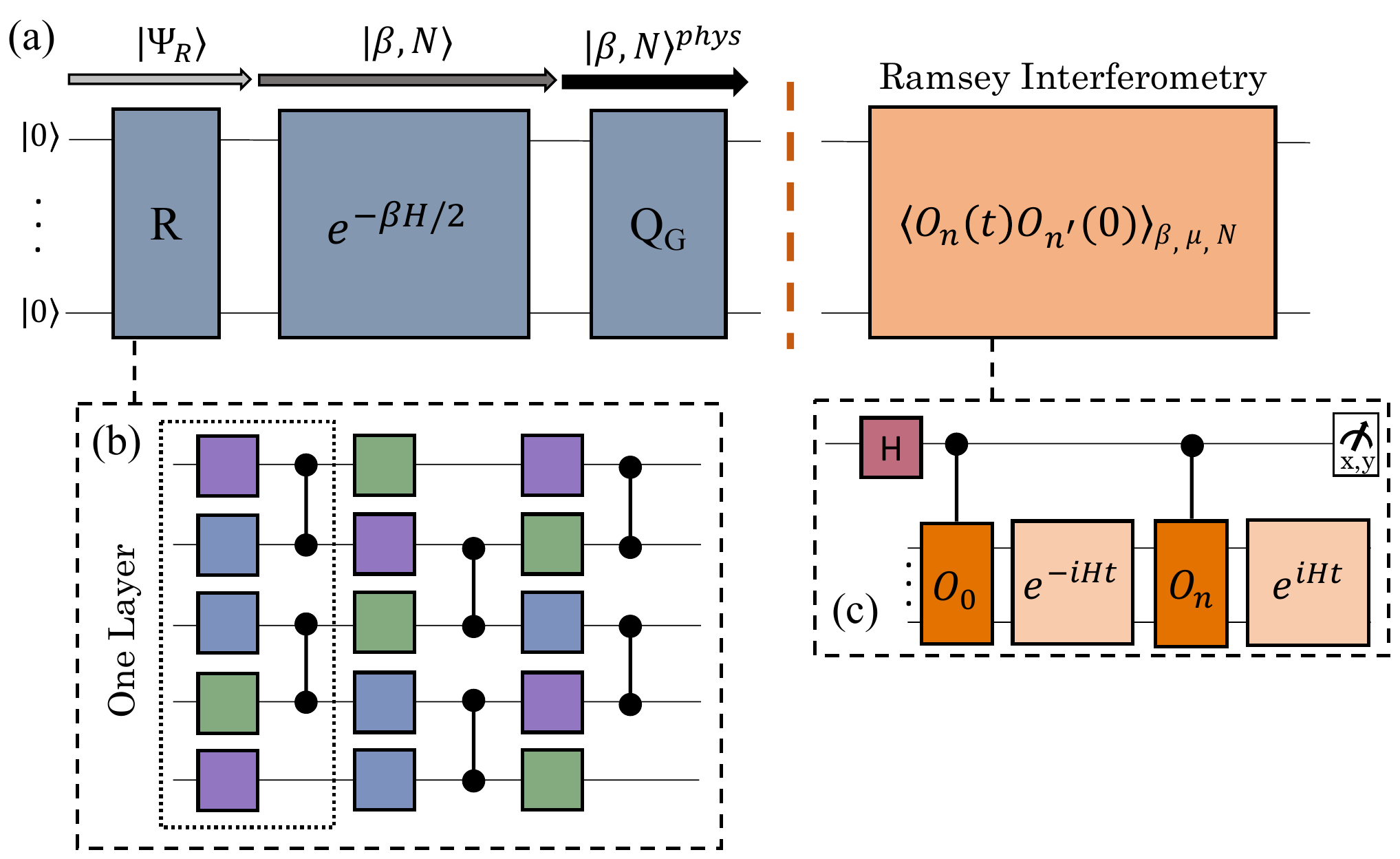}
    \caption{(a) PTPQ-state preparation circuit. (b) Random-state preparation sub-circuit structure, see Ref.~\cite{richter2021simulating} for details. (c) Ramsey interferometry sub-circuit required for the calculation of non-equal time correlation functions. $\mathsf{H}$ is a Hadamard gate. Note that the final time-evolution block in the interferometry circuit can be replaced by a $Z$ rotation of the ancilla qubit for further simplification~\cite{wecker2015solving}.}
    \label{fig:circuitoverview}
\end{figure}

A high-level overview of a circuit for preparing PTPQ states on quantum computers is summarized in \Fig{fig:circuitoverview}(a), including a standard sub-circuit in \Fig{fig:circuitoverview}(b) to efficiently prepare a (pseudo-) Haar-random state~\cite{richter2021simulating}. The system is imaginary-time evolved with the application of $e^{-\beta \widetilde{H}/2}$. Since $[G_n,H]=0$, one can separate the imaginary-time evolution into $e^{-\beta H/2}$, followed by Gauss's law enforcing $Q_G \equiv e^{-\beta \sum_n f(G_n)/2}$. 
Finally, thermal observables can be measured as prescribed by \Eq{eq:TPQobs}, and a Ramsey interferometry circuit (orange), shown in \Fig{fig:circuitoverview}(c), allows the computation of thermal non-equal time correlation functions~\cite{somma2002simulating}.

\section{Thermal phase diagram of $\mathbb{Z}_2^{1+1}$}
\noindent
To illustrate the approach and without loss of generality, we focus on a simple prototype model, $\mathbb{Z}_2$ LGT in $1+1$ spacetime dimensions ($\mathbb{Z}_2^{1+1}$), a case where classical simulations allow benchmarking the algorithm for small systems. The $\mathbb{Z}_2^{1+1}$ Hamilltonian is
\begin{equation}
\begin{aligned}\label{eq:full_Z2}
    H=  \frac{1}{2a}\sum_{n=0}^{
    N-2}(c^\dagger_{\scriptsize n} \Tilde{\sigma}^z_{\scriptsize n} c_{\scriptsize n+1} +{\rm H.c.})+m\sum_{n=0}^{N-1} (-1)^n c^\dagger_n c_n - \epsilon \sum_{n=0}^{N-2} \Tilde{\sigma}^x_n\,,
\end{aligned}
\end{equation}
where $c^\dagger_n$ ($c_n$) are fermionic creation (annihilation) operators, and $\tilde{\sigma}^z_n$ and $\tilde{\sigma}^x_n$ are Pauli spin operators realizing the $\mathbb{Z}_2$ link and electric-field operators, respectively. $N$, $a$, $m$, and $\epsilon$ are lattice size and spacing, fermion mass, and electric-field coupling, respectively. Gauss's law operator is given by $G_n\equiv \Tilde{\sigma}^x_n \Tilde{\sigma}^x_{n-1}e^{i \pi \mathcal{Q}_n}$ where $\mathcal{Q}_n \equiv  c^\dagger_n c_n + [(-1)^n -1]/{2}$ is the fermion charge. We work with open boundary conditions. Gauss's law is enforced by adding $\sum_nf(G_n)=\lambda \sum_n (1-G_n)$ to the Hamiltonian, where $\lambda$ is taken to be large compared with other mass scales in the problem.

\begin{figure}[t]
\begin{centering}
\includegraphics[width=0.5\textwidth]{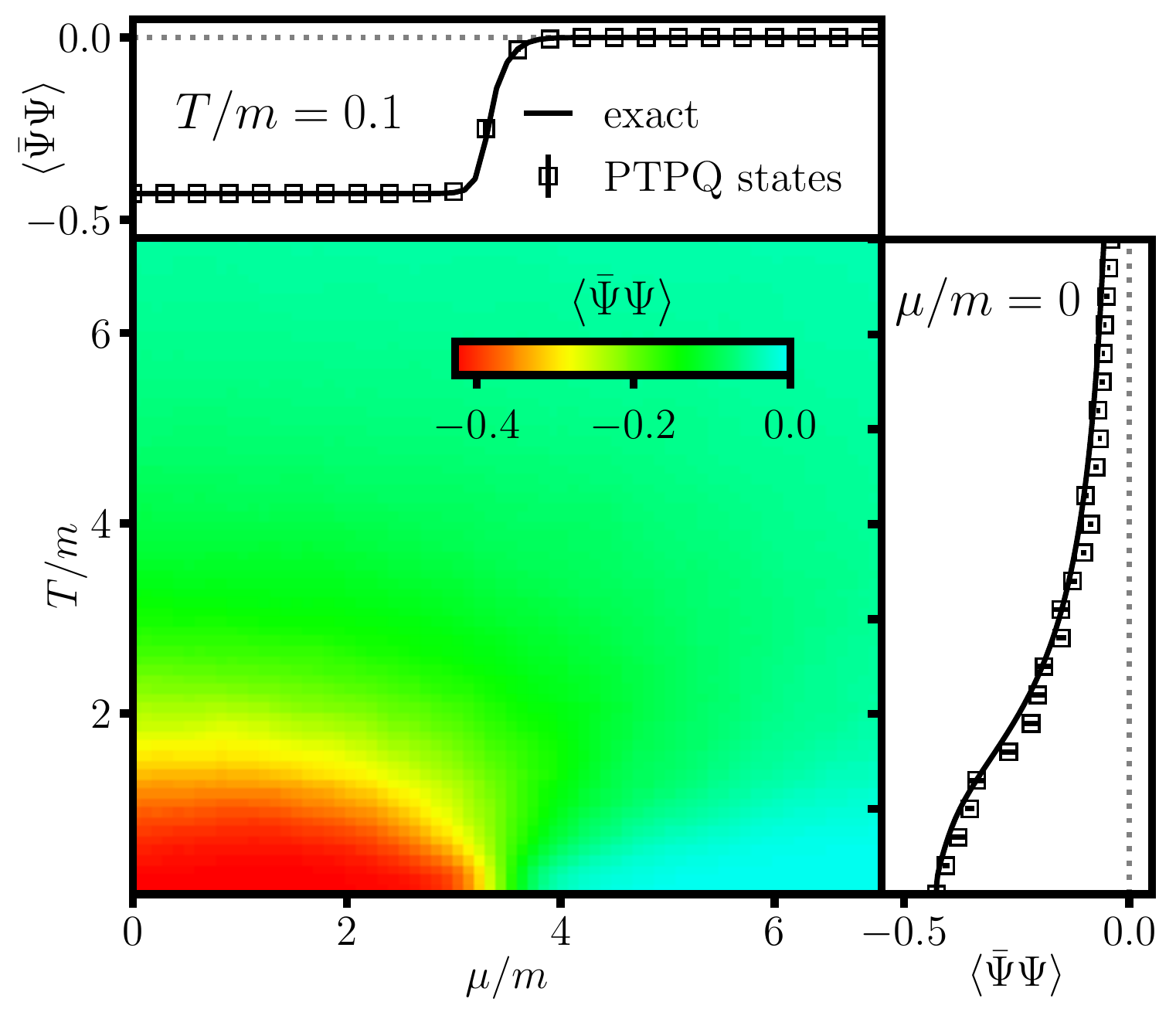}
\caption{Chiral phase diagram for $\mathbb{Z}_2^{1+1}$ LGT with staggered fermions with $N=6$ and $\lambda/m=4$, calculated with PTPQ states with random-circuit depth $d=20$. Results are averaged over $r=10$ PTPQ-state realizations, and error bars denote statistical uncertainty. Figure is adopted from Ref.~\cite{davoudi2022toward}.}
\label{fig:chiral_plots}
\end{centering}
\end{figure}

Figure~\ref{fig:chiral_plots} shows the chiral phase diagram of the model for $N=6$ lattice sites. We plot the chiral condensate $\langle \Bar{\Psi}\Psi\rangle \equiv \frac{1}{N}\langle\sum_{n=0}^{N-1}(-1)^n c^\dagger_n c_n \rangle$, computed
with PTPQ states. A Haar-random circuit depth of $d=20$ is used, along with the Gauss's law penalty coefficient $\lambda/m=4$, and results are obtained from an average over 
$r=10$ PTPQ realizations. Similar to QCD,  chiral symmetry is broken at low temperatures and chemical potential, while at large temperatures and densities it is restored. Side panels show the zero temperature and density limits, respectively, along with comparison with an exact computation (solid lines). A sharp transition is anticipated at $T=0$ (in the thermodynamic limit), while the finite-temperature transition is displaying crossover behavior.

The approach also allows the computation of non-equal time correlation functions of thermal states, relevant e.g., to estimate transport quantities of the Quark-Gluon-Plasma (QGP) created in relativistic heavy-ion collisions. Because they involve matrix elements of thermal states separated in \textit{real time}, even at zero quark chemical potential they are difficult to compute using Monte-Carlo methods which are restricted to euclidean spacetime. In contrast, real-time evolution is most natural for quantum computers. Using the Ramsey interferometry scheme shown in \Fig{fig:circuitoverview}(c),  one can realize the quantum-mechanical superposition of states separated in real time through the use of an ancilla qubit~~\cite{somma2002simulating}.

\begin{figure}[t]
\begin{centering}
\includegraphics[width=0.5\textwidth]{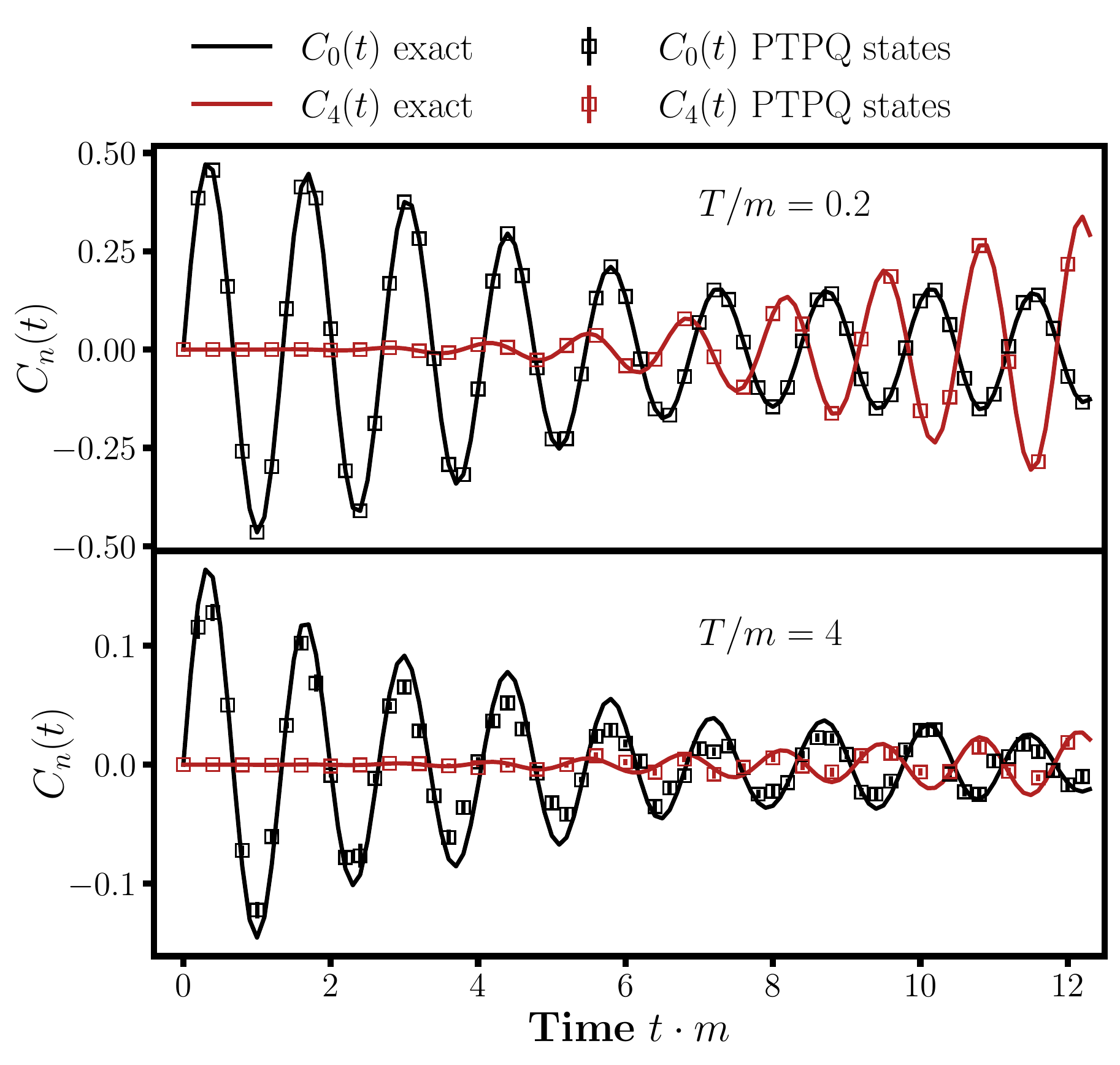}
\caption{Non-equal time correlation functions $C_0(t)$ and $C_4(t)$ at $N=6$ fermionic sites, $\mu/m=0.5$, and $\lambda/m=4$. Results from a single PTPQ state are shown for $T/m=0.2$, and data from $r=20$ PTPQ states are averaged over at $T/m=4$. Figure is adopted from Ref.~\cite{davoudi2022toward}.}
\label{fig:nonequal_time_correlators}
\end{centering}
\end{figure}

As an example, in \Fig{fig:nonequal_time_correlators} we present results of the thermal correlator $C_n(t) \equiv \langle [j_n(t), j_0(0) ]\rangle_{\beta}$ with $j_n(t) \equiv e^{iHt} j_ne^{-iHt}$ and $j_n \equiv \frac{i}{2a}[\sigma^+_n \tilde{\sigma}^z_n \sigma^-_{n+1} - \rm{H.c.}]$ evaluated using PTPQ states for two different temperatures. The PTPQ-states results (symbols) reproduce the exact results (solid lines) well, even with a single $r=1$ realization for $T/m=0.2$ and $r=20$ realizations for $T/m=0.4$, consistent with expectations that the statistical error should increase with temperature~\cite{sugiura2013canonical}. To reach the zero-frequency limit ($\omega \rightarrow 0$), to compute e.g., the thermal conductivity via $\sigma_{\beta,\mu} \sim \lim\limits_{\omega\rightarrow 0 } {a}/{\omega} \int dt \, e^{i\omega t} \sum_n C_n(t)$, significantly larger lattices are required than achievable by classical emulation of the quantum device. This makes even this simple model an attractive 
target for near-future quantum computation.
%

\section{Considerations for NISQ era and far-term circuit implementation}
\begin{figure}[ht]
\begin{centering}
\includegraphics[width=0.99\textwidth]{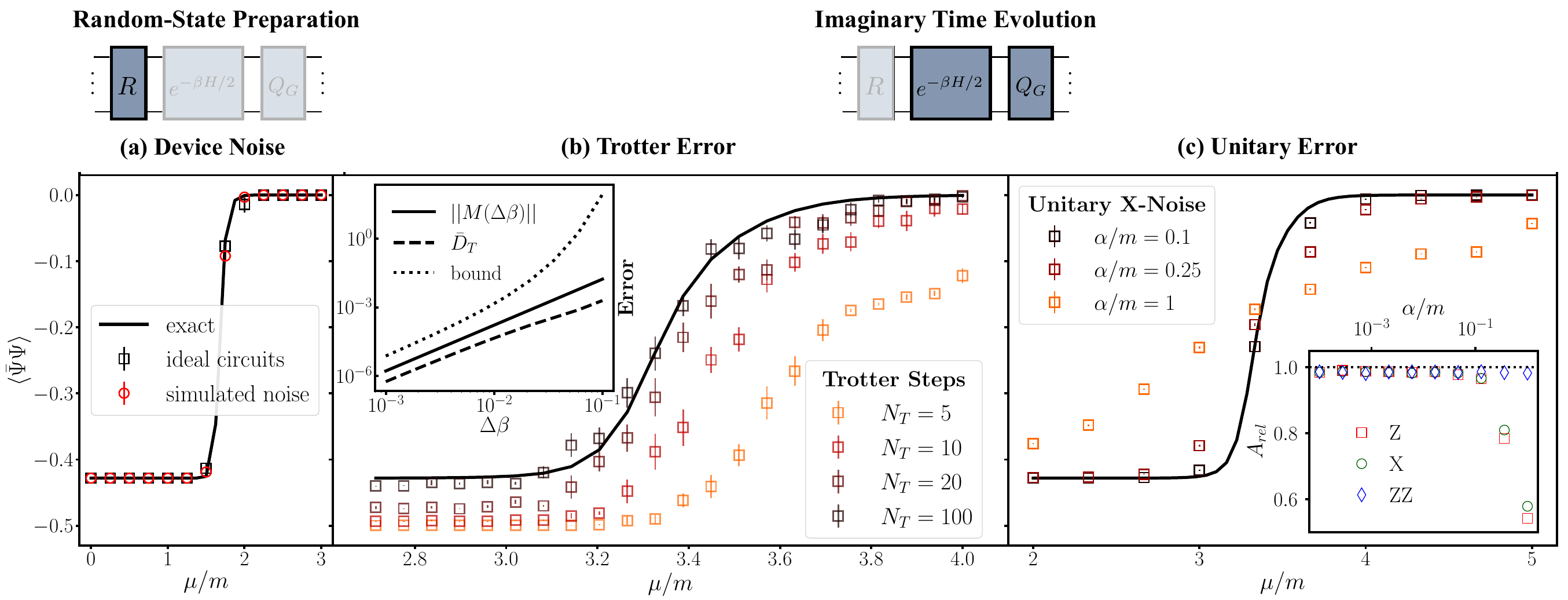} 
\end{centering}
\caption{(a) Chiral condensate $\langle \bar{\Psi} {\Psi} \rangle$ as a function of $\mu/m$ with simulated device noise ~\cite{Qiskit} applied to the random-state preparation sub-circuit. (b) Effects of Trotter error. Inset: Multiplicative Trotter error $||M(\Delta \beta)||$ using first-order product formulas~\cite{childs2021theory}, as well as the mean trace distance $\bar{D}_{Tr}$ between $e^{-\Delta \beta H}|\Psi_R\rangle$ and $S(\Delta \beta)|\Psi_R\rangle$, are plotted as functions of imaginary time-step size $\Delta \beta$ at $\mu/m=4$.  (c) Impacts of bit-flip (X) effective Hamiltonian errors  at various relative error strengths $\alpha/m$. Inset: Relative accuracy $A_{\rm rel}$ as a function of $\alpha/m$ with various error types at $\mu/m=3$. All results are with $d=20$, $r=10$, $\lambda/m=8$,$N=6$, $T/m=0.1$, and  averaged over $r=50$ random error terms with strength $\alpha/m$. Figure is adopted from Ref.~\cite{davoudi2022toward}.}
\label{fig:error_plots}
\end{figure}

\noindent
Noisy intermediate-scale quantum (NISQ) devices involve limited algorithmic complexity and are prone to a multitude of errors, depending on device type and the problem investigated. The PTPQ-state computation of thermal properties of a LGT, exemplified in this talk,
consists of two basic components: (pseudo-) Haar-random-state preparation, tried and proven
with existing hardware, as well as non-unitary imaginary-time evolution.
Figure~\ref{fig:error_plots}(a) illustrates what may be expected from (pseudo-) Haar-random-state preparation on an actual device, here a simulator of IBM's $\tt{ibmq\_sydney}$ device via Qiskit~\cite{Qiskit}, using standard circuits~\cite{richter2021simulating} but with (classically emulated) exact imaginary-time evolution. The results are seen to be robust to these errors.

In contrast, imaginary-time evolution is a less established and more challenging aspect of this proposal because quantum computers do not directly provide non-unitary operations. Nevertheless, techniques for non-unitary operations are anticipated to be important in physics problems and are under intense investigation. An important example is ground- or excited-state preparation where one may find advantage over e.g., adiabatic state preparation or variational algorithms~\cite{motta2020determining}. At present, there exists no consensus as to the optimal imaginary-time evolution algorithm, with proposed algorithms either involving approximate evolution and classical optimization or ancilla degrees of freedom. Therefore, their capabilities in terms of scalability and efficiency for near- versus far-term implementation vary widely, see e.g., Refs.~\cite{motta2020determining,low2019hamiltonian,camps2022fable}. 

Common to most imaginary-time evolution algorithms, like in unitary real-time evolution, is Trotterization of the evolution operator $S(\beta) = e^{-\beta H/2}$. In the following, we will work with first-order product formulas,
\begin{align}
\label{eq:Trotterform}
    S(\beta) \equiv [S(\Delta \beta)]^{N_T} \equiv \Big[ \prod_{\gamma=1}^\Gamma e^{- H_{\gamma} \Delta \beta} \Big]^{N_T}\,,
\end{align}
where $N_T = \beta/ (2\Delta \beta)$ is the number of Trotter steps of length $\Delta\beta$ of the interval $\beta/2$, $H_\gamma$ are $\Gamma$ (local) components of the Hamiltonian $H$. Since no Trotter error is expected from separately implementing the penalty term $Q_G$, this term is not considered here. Errors from Trotterization are independent of device performance.

Figure~\ref{fig:error_plots}(b) summarizes the effects of Trotterization, estimated by varying the number of Trotter time steps for finite-temperature chiral-condensate calculations at $T/m=0.1$. For percent-level agreement, $N_T\gtrsim 100$ is required, while at $N_T=10$, the chiral phase transition is qualitatively visible. The inset contains the multiplicative Trotter error as a function of imaginary-time step size, $M(\Delta \beta) \equiv S(\Delta \beta)/e^{-\Delta \beta \, H}-I$, and the mean trace distance $\bar{D}_{Tr}$ between $e^{-\Delta \beta H}|\Psi_R\rangle$ and $S(\Delta \beta)|\Psi_R\rangle$ averaged over $r=10$ PTPQ states. The analytical bound on the multiplicative Trotter error obtained in Ref.~\cite{childs2021theory} is also plotted. Next, motivated by phase-flip, bit-flip, and cross-talk errors commonly seen in NISQ-era devices, and to estimate effects of device miscalibration and errors with minimal algorithmic bias, we consider the so-called unitary errors, parameterized by an effective Hamiltonian $H'=\widetilde{H}+H_{\rm err}^\alpha$, where $H_{\rm err}^\alpha$ represents 1- or 2-local errors applied to each qubit, with randomized weights bounded by a strength parameter $\alpha$. Our error model is given by $H_{\rm err} = \sum_{l=0}^{2N-1} K_l^{\alpha} \sigma_l$, where $\sigma_l \in\{ \sigma^x_l,\sigma^z_l, \sigma^z_l\sigma^z_{l+1}  \}$ and $K_l^\alpha$ is drawn uniformly from $[-\alpha,\alpha]$. There is no distinction between fermionic sites or gauge links because the errors apply to all qubits equally. The results of this study for bit-flip (X) errors are shown in \Fig{fig:error_plots}(c), where we display the chiral condensate obtained for varying strengths of $\alpha$. The inset shows the relative accuracy $A_{\rm rel} \equiv 1-{| \langle  \bar{\Psi} {\Psi} \rangle^{PTPQ}(\alpha) - \langle  \bar{\Psi} {\Psi} \rangle|}/{| \langle  \bar{\Psi} {\Psi} \rangle|}$ for phase (Z), bit-flip (Z), and correlated phase errors (ZZ) as a function of $\alpha/m$.

Finally, we discuss scaling and resource requirements of the proposal. It is sufficient to prepare pseudo-Haar-random states (as opposed to true Haar-random states), which is possible with relatively shallow, polynomial-depth circuits and will not be discussed further. To estimate the imaginary-time evolution cost, one may consider the QITE algorithm~\cite{motta2020determining}, based on approximating each non-unitary evolution component in \Eq{eq:Trotterform} with a unitary evolution of similar or greater non-locality, using a classical optimization procedure. This approximation introduces another source of error, in addition to the Trotter error. While a more detailed analysis can be found in Refs.~\cite{motta2020determining,davoudi2022toward}, we just note in summary that the QITE circuit complexity is exponential in the correlation length of the state acted on.  While the QITE algorithm starts at an infinite-temperature state with zero correlation length, the cost of QITE may become inhibiting e.g., at phase transitions, illustrating the importance of developing more efficient non-unitary evolution quantum algorithms.

\section{Conclusion}
\noindent
In this talk, we discussed a novel approach toward quantum computing finite-temperature phase diagrams of lattice gauge theories, to eventually overcome the infamous sign-problem of finite-density QCD, with important implications for the understanding of the quark-gluon plasma in relativistic heavy-ion collision,
e.g., in the search for the QCD critical point,
or for elucidating the composition the interior of neutron stars. 
Another important goal is to quantify QCD transport through computation of thermal non-equal time correlation functions.  While quantum computing the QCD phase diagram is still a far-term endeavor, we demonstrated how progress can be made using a simple but relevant prototype model to compute the thermal phase diagram as well as non-equal time correlation functions by developing relevant quantum algorithms. Our approach is another motivation for the development of efficient non-unitary evolution algorithms on quantum computers, an important subject generating significant interest across different fields. We expect that quantum information sciences will make important contributions, conceptually and computationally, to understanding the structure of strongly coupled gauge theories at finite temperature and chemical potential, with significant interdisciplinary synergies expected e.g., in quantum thermodynamics and condensed-matter physics.  

 
\section*{Acknowledgements}
Z.D. and N.M. acknowledge funding by the U.S. Department of Energy’s Office of Science, Office of Nuclear Physics under Award no. DE-SC0021143. Z.D and C.P. acknowledge funding by the U.S. Department of Energy’s Office of Science, Office of Advanced Scientific Computing Research, Accelerated Research in Quantum Computing program award DE-SC0020312. N.M. further acknowledges funding by the U.S. Department of Energy, Office of Science, Office of Nuclear Physics, InQubator for Quantum Simulation (IQuS) (\url{https://iqus.uw.edu}) under Award Number DOE (NP) Award DE-SC0020970. C.P. was further supported by the National Science Foundation under QLCI grant OMA-2120757.
\bibliographystyle{JHEP}
\bibliography{references_cleaned}


\end{document}